\documentclass{eas}
\usepackage{graphicx}
\begin{document}

\TitreGlobal{Mass Profiles and Shapes of Cosmological Structures}

\title{The OSER project}

\author{Moniez, M.}\address{Laboratoire de l'Acc\'{e}l\'{e}rateur Lin\'{e}aire,
{\sc IN2P3-CNRS}, Universit\'e de Paris-Sud, \\
B.P. 34, 91898 Orsay Cedex, France.
E-mail: moniez@lal.in2p3.fr}
\runningtitle{The OSER project}
\setcounter{page}{1}
\index{Moniez, M.}
%
\begin{abstract}
The OSER project (Optical Scintillation by Extraterrestrial Refractors)
is proposed to search for scintillation of extragalactic sources
through the galactic -- disk or halo -- transparent $\mathrm{H_2}$ clouds,
the last unknown baryonic structures.
This project should allow one to detect column density
stochastic variations in cool Galactic molecular clouds of order of
$\sim 3\times 10^{-5}\,\mathrm{g/cm^2}$
per $\sim 10\,000\,\mathrm{km}$ transverse distance.
\end{abstract}
\maketitle
%
\section{The transparent baryonic matter and its possible signature}
\begin{figure}[h]
\centering
\vbox{
\includegraphics[width=6cm]{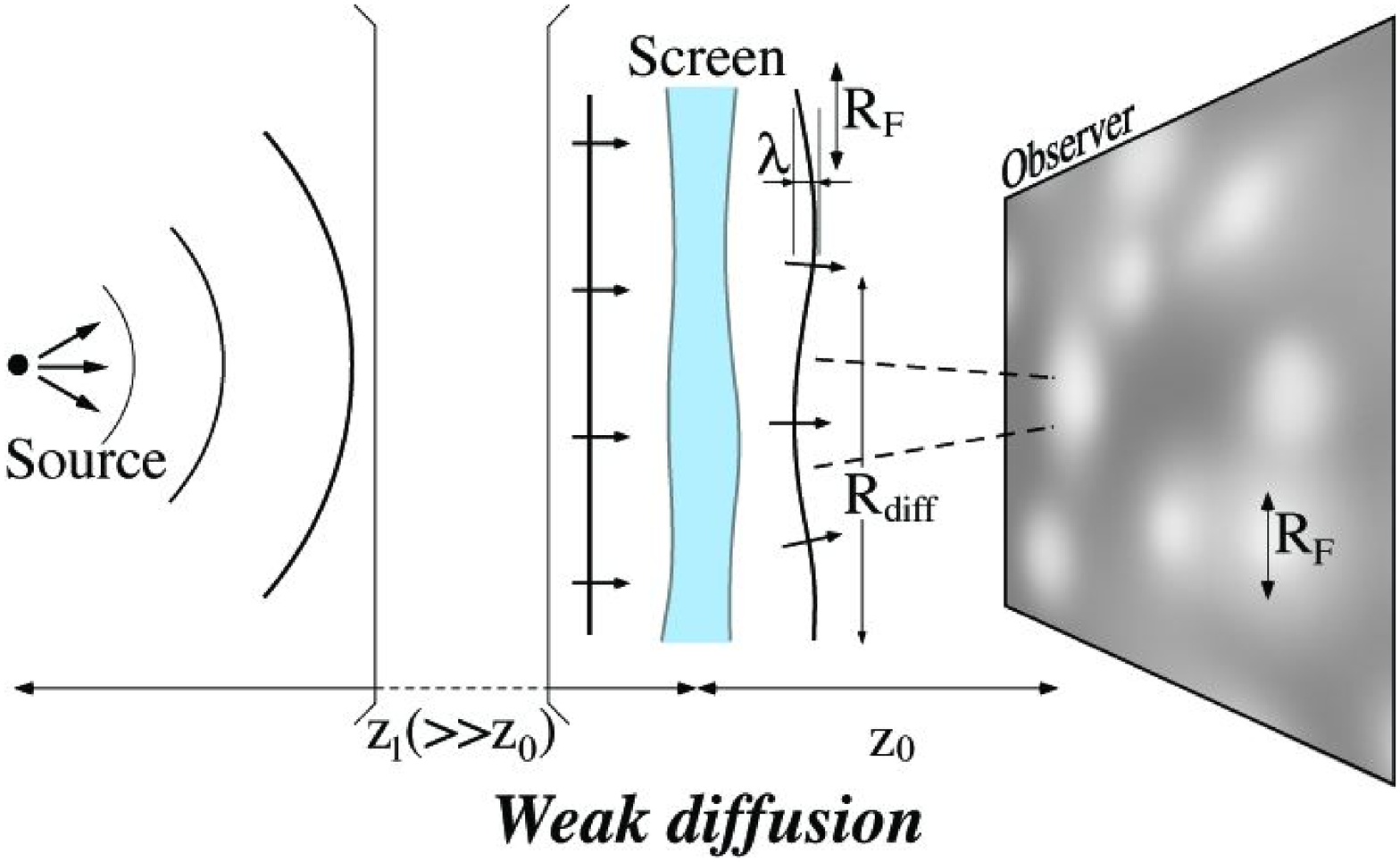}
\includegraphics[width=6cm]{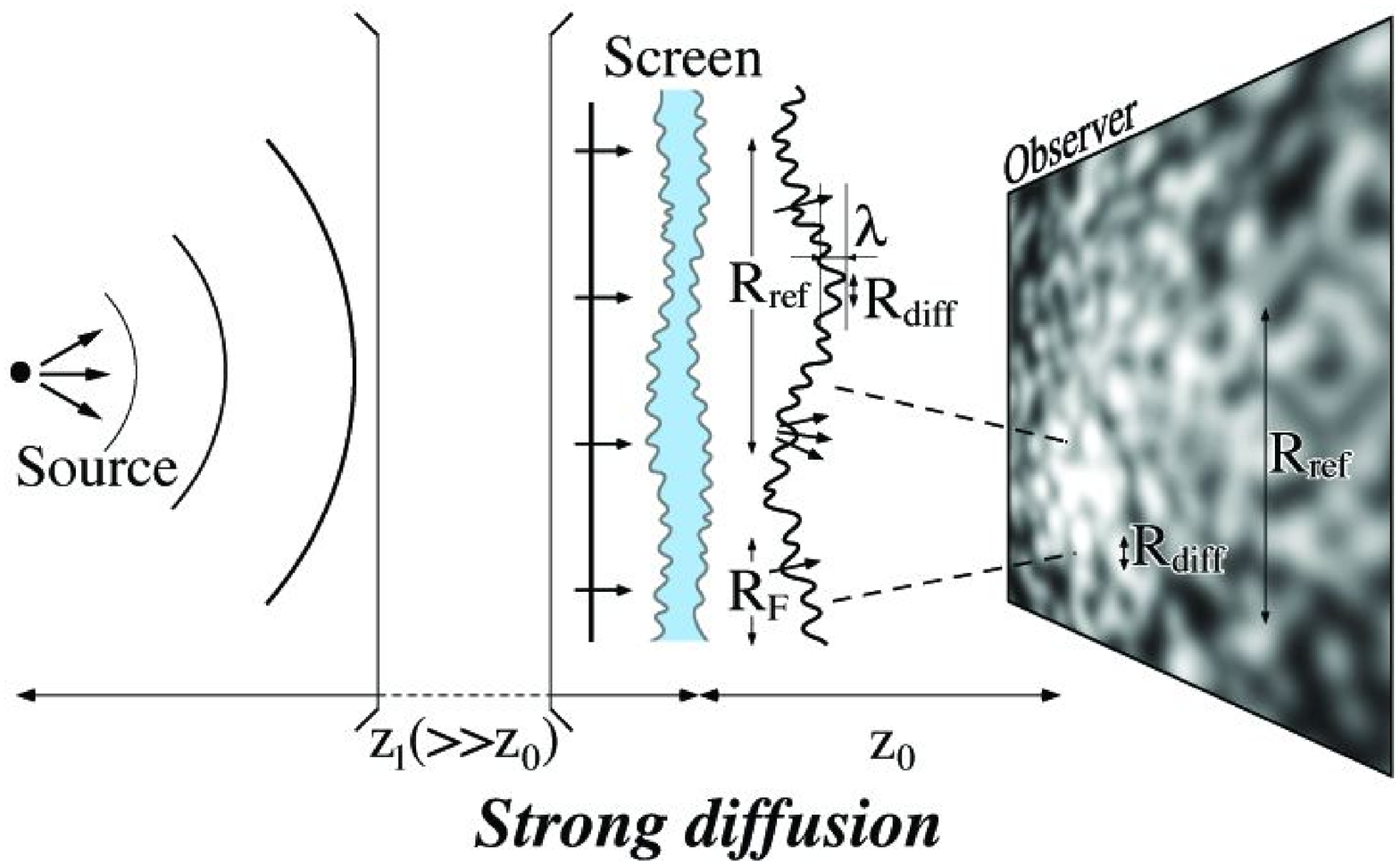}
}
\caption[]{\it The two scintillation regimes (see text):
$R_{diff}\gg R_F$ (left), 
$R_{diff}< R_F$ (right). 
}
\label{front}
\end{figure}

Considering the results of baryonic compact massive objects searches
(Lasserre {\it et al.} \cite{notenoughmachos}; \cite{SMC5ans}; \cite{MACHO}),
cool molecular hydrogen ($\mathrm{H_2}$) clouds
should now be seriously considered as a possible major component
of the Galactic baryonic hidden matter.
It has been suggested that a hierarchical structure of cold $\mathrm{H_2}$
could fill the Galactic thick disk (\cite{fractal}) or halo (\cite{Jetzer1}), 
providing a solution for the Galactic hidden matter problem.
This gas should form transparent
``clumpuscules'' of $10\,\mathrm{AU}$ size,
with a column density of $10^{24-25}\,\mathrm{cm^{-2}}$, and a
surface filling factor smaller than 1\%.
Refraction through such
an inhomogeneous transparent $\mathrm{H_2}$ cloud (hereafter called screen)
distorts the wave-front of incident electromagnetic waves
(Fig. \ref{front}, see \cite{Moniez} for details).
The extra optical path induced by a 
screen at distance $z_0$ can be described 
by a function $\delta(x_1,y_1)$
in the plane transverse to the observer-source line.
The amplitude in the observer's plane after propagation
is described by the Huygens-Fresnel diffraction theory:
\begin{equation}
\label{amplit}
A_0(x_0,y_0)=\frac{Ae^{2i\pi z_0/\lambda}}{2i\pi R_F^2}
\times \int\!\!\int_{-\infty}^{+\infty}
e^{\frac{2i\pi\delta(x_1,y_1)}{\lambda}}
e^{i\frac{(x_0-x_1)^2+(y_0-y_1)^2}{2 R_F^2}}dx_1dy_1 ,
\end{equation}
where $A$ is the incident amplitude (before the screen),
taken as a constant for a very distant on-axis point-source,
and $R_F=\sqrt{\lambda z_0/2\pi}$ is the Fresnel radius.
$R_F$ is of order of $1500\,\mathrm{km}$ to $15\,000\,\mathrm{km}$
at $\lambda=500\,\mathrm{nm}$,
for a screen distance $1\,\mathrm{kpc}<z_0<100\,\mathrm{kpc}$.
$R_F$ characterizes the $(x_1,y_1)$ domain that contributes
significantly to the integral (a few Fresnel radii).
For a {\it point-like} source, the intensity
in the observer's plane shows interferences (speckle) if 
$\delta(x_1,y_1)$ varies stochastically of order of $\lambda$
within the Fresnel radius domain.
This variation rate is comparable to
the average gradient that characterizes the hypothetic
$\mathrm{H_2}$ structures.

\begin{figure}[h]
\centering
\mbox{
\includegraphics[width=4cm]{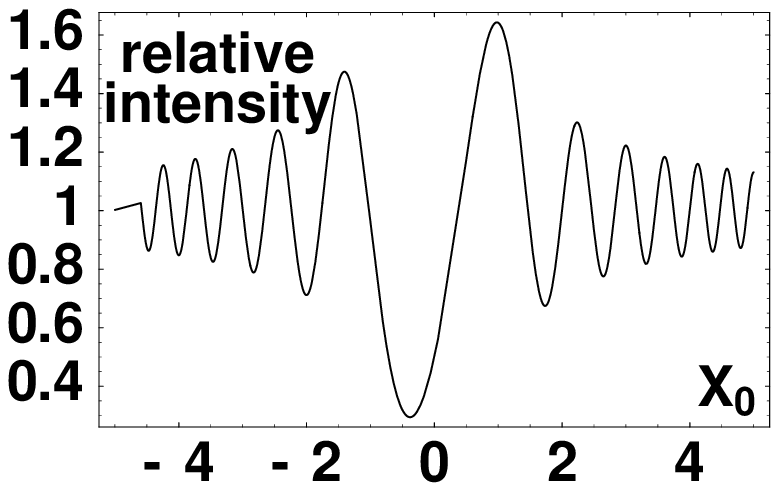}
\includegraphics[width=4cm]{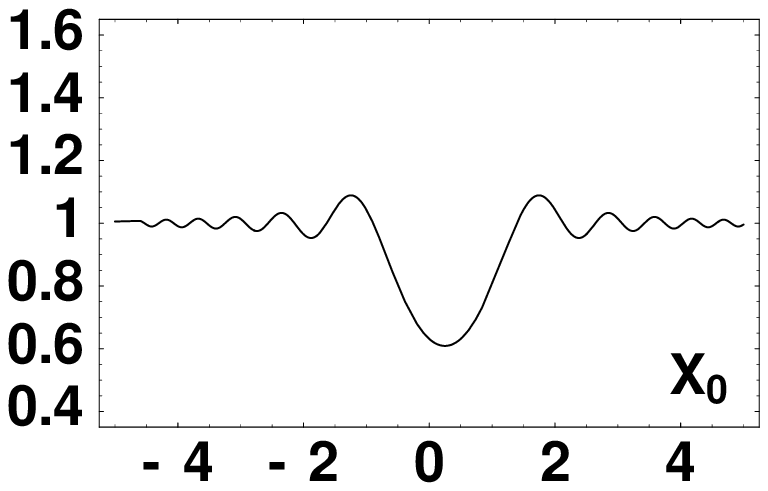}
\includegraphics[width=4cm]{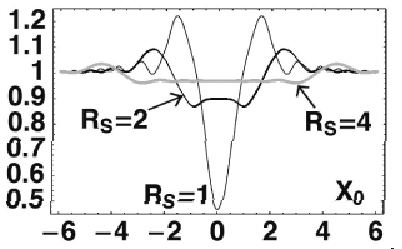}
}
\caption[]{\it 
diffraction patterns in the
observer's plane (see text).
%
%
%
The $X_0-\mathrm{axis}$ ($X_0=x_0/\sqrt{\pi}R_F$)
is perpendicular to the structure's edge.
The origin is the intercept
of the source-step line with the observer's plane.
\label{diffpoint}}
\end{figure}

As for radio-astronomy (\cite{lyne}),
the stochastic variations of $\delta(x_1,y_1)$ are characterized
by the diffusion radius $R_{diff}$, defined as the
transverse separation
for which the root mean square of the optical path difference
is $\lambda/2\pi$.\\
- If $R_{diff}\gg R_F$, the screen is weakly diffusive and moderately
distorts the wavefront, producing
low contrast patterns with length scale $R_F$ in the observer's plane
(Fig. \ref{front} left).\\
- If $R_{diff}< R_F$, the screen is strongly diffusive;
two modes occur, the diffractive scintillation -- producing strongly
contrasted patterns characterized by the length scale $R_{diff}$ --
and the refractive scintillation -- giving less contrasted patterns and
characterized by the large scale structures of the screen $R_{ref}$ --
(Fig. \ref{front} right).\\
{\bf Basic configurations:}
\label{simplecase}
Fig. \ref{diffpoint} (left and center) displays the expected intensity
variations in the observer's plane
for a {\it point-like} monochromatic source observed through a transparent screen
with a step of optical path $\delta=\lambda/4$ and through a prism edge.
The inter-fringe is $\sqrt{\pi}R_F$, the unique distance scale here.
\section{Limitations from spatial and temporal coherences}
At optical wavelengths, the diffraction pattern contrast
is severely limited by the size of the incoherent source $r_s$.
Fig. \ref{diffpoint} (right) shows the diffraction
patterns for various reduced source radii, defined as
$R_S=r_s/(\sqrt{\pi}R_F)\times z_0/z_1$, where $z_1$ is the distance
from the source to the screen.
In return, temporal coherence with the standard UBVRI filters
is sufficiently high to enable the formation of contrasted interferences
in the configurations considered here.
\section{What is to see?}
An interference pattern with inter-fringe of
$\sim R_F$ ($1000-10\,000\,\mathrm{km}$ at $\lambda=500\,\mathrm{nm}$)
is expected to sweep across the Earth when
the line of sight of a sufficiently small astrophysical source
crosses an inhomogeneous transparent Galactic
structure (Table \ref{configurations}).
This pattern moves at the relative transverse velocity $V_T$ of the screen.
In the present paper, we assume that the scintillation is
mainly due to pattern motion rather than pattern instability
(frozen screen hypothesis),
as it is usually the case in radioastronomy observations (\cite{lyne}).
At the distance of the Galactic $\mathrm{H_2}$ clouds we are interested in,
we expect a typical modulation index $m_{scint}$ (or contrast)
ranging between 1\% and $\sim 20\%$ at $\lambda=500 nm$,
critically depending on the source apparent size;
the time scale $t_{scint} = R_F/V_T$ of the intensity variations
is of order of a minute.
As the inter-fringe scales with $\sqrt{\lambda}$,
one expects a significant difference in the time scale $t_{scint}$
between the red side of the optical
spectrum and the blue side. This property
might be used to sign the diffraction
phenomenon at the $R_F$ natural scale.
\begin{table*}
\begin{center}
\setlength{\tabcolsep}{2pt}
{\small
\caption[]{\it Configurations leading to strong diffractive scintillation
assuming $R_{diff}\le R_F$.
Numbers are given for $\lambda=500\,\mathrm{nm}$.
}
\label{configurations}
{\footnotesize
\begin{tabular}{|c|c||c|c|c||c|c|c|}
\cline{3-8}
\multicolumn{2}{c||}{} & \multicolumn{6}{c|}{\bf SCREEN} \\
\cline{3-8}
\multicolumn{2}{c||}{} & atmos- & solar	& solar	& \multicolumn{3}{c|}{Galactic} \\
\multicolumn{2}{c||}{} & phere & system& suburbs  & thin disk& thick disk& halo \\
\hline
\multicolumn{2}{|r||}{Distance}	& $10\,\mathrm{km}$	& $1\,\mathrm{AU}$ & $10\,\mathrm{pc}$	& $300\,\mathrm{pc}$ & $1\,\mathrm{kpc}$ & $10\,\mathrm{kpc}$ \\
\hline
\multicolumn{2}{|r||}{$R_F$ to $\times$ by $\left[\frac{\lambda}{500\,\mathrm{nm}}\right]^{\frac{1}{2}}$} & $2.8\,\mathrm{cm}$	& $109\,\mathrm{m}$& $157\,\mathrm{km}$	& $860\,\mathrm{km}$& $1570\,\mathrm{km}$ & $5000\,\mathrm{km}$ \\
\hline
\multicolumn{2}{|r||}{Transverse speed $V_T$}	& $1\,\mathrm{m/s}$	& $10\,\mathrm{km/s}$ & $20\,\mathrm{km/s}$	& $30\,\mathrm{km/s}$ & $40\,\mathrm{km/s}$ & $200\,\mathrm{km/s}$ \\
\hline
\multicolumn{2}{|r||}{$t_{scint}$
to $\times\left[\frac{\lambda}{500\,\mathrm{nm}}\right]^{\frac{1}{2}}
\left[\frac{R_{diff}}{R_F}\right]$}	& $0.03\,\mathrm{s}$ & $0.01\,\mathrm{s}$ & $8\,\mathrm{s}$ & $29\,\mathrm{s}$ & $40\,\mathrm{s}$ & $25\,\mathrm{s}$ \\
\hline
\multicolumn{2}{|r||}{Optical depth $\tau_{scint}$} & $1$ & & & \multicolumn{3}{c|}{total $>10^{-7}$} \\
\hline
\multicolumn{2}{|r||}{$m_{scint}$ in \% to multiply by} & & $32\%\times$ & $2.2\%\times$ & $4.1\%\times$ & $2.2\%\times$ & $7.1\%\times$ \\
\multicolumn{2}{|r||}{
$\left[\frac{\lambda}{500\,\mathrm{nm}}\right]^{\frac{1}{2}}
\left[\frac{R_{diff}}{R_F}\right]\left[\frac{r_s}{r_{\odot}}\right]^{-1}$
}	& $100\%$	& $\left[\frac{d}{10\mathrm{pc}}\right]$ & $\left[\frac{d}{1\mathrm{kpc}}\right]$	& $\left[\frac{d}{10\mathrm{kpc}}\right]$	& $\left[\frac{d}{10\mathrm{kpc}}\right]$ & $\left[\frac{d}{100\mathrm{kpc}}\right]$ \\
\hline
\hline
\multicolumn{2}{|c||}{\bf SOURCE} & \multicolumn{6}{c|}{\bf DIFFRACTIVE MODULATION INDEX $m_{scint}$} \\
\cline{1-2}
Location & Type & \multicolumn{6}{c|}{\bf (to multiply by $\sqrt{\lambda/500\,\mathrm{nm}}\times R_{diff}/R_F$)} \\
\hline
\cline{3-8}
nearby 10pc	& any star	& $\ll 1\%$ & $<100\%$ & & & & \\
\cline{1-2}\cline{4-8}
Galactic 8kpc & star & in a & 100\% & 1-70\% & 1-10\% &  &  \\
\cline{1-2}\cline{4-8}
LMC 55kpc & A5V ($r_s=1.7r_{\odot}$) & tele- & 100\% & 70\% & 13\% & 7\% & 2\% \\
\cline{1-2}\cline{4-8}
M31 725kpc & B0V ($r_s=7.4r_{\odot}$) & scope & 100\% & 100\% & 40\% & 22\% & 7\% \\
\cline{1-2}\cline{4-8}
z=0.2--0.9Gpc	& SNIa & $>1\,\mathrm{m}$ & 100\% & 70\% & 13\% & 7\% & 2\%\\
\cline{1-2}\cline{4-8}
z=1.7--1.7Gpc& Q2237+0305 & & 100\% & $>45\%$ & $>8\%$ & $>4\%$ & $>1.4\%$\\
\hline\end{tabular}
}
}
\end{center}
\end{table*}
\section{Feasibility studies: simulation}
\begin{figure}[h]
\centering
\parbox{6cm}{
\includegraphics[width=6cm]{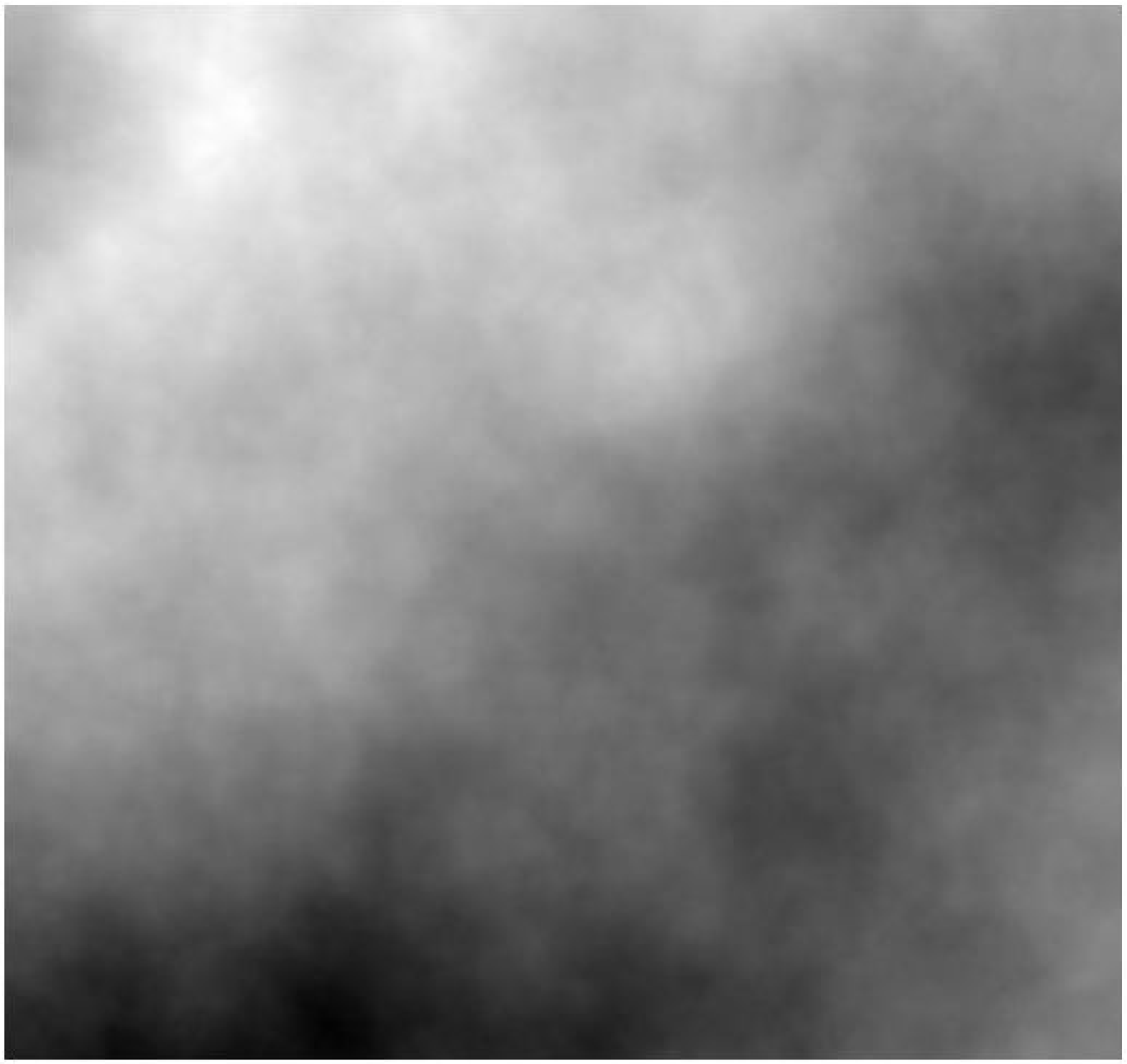}}
\parbox{6cm}{
\includegraphics[width=6cm]{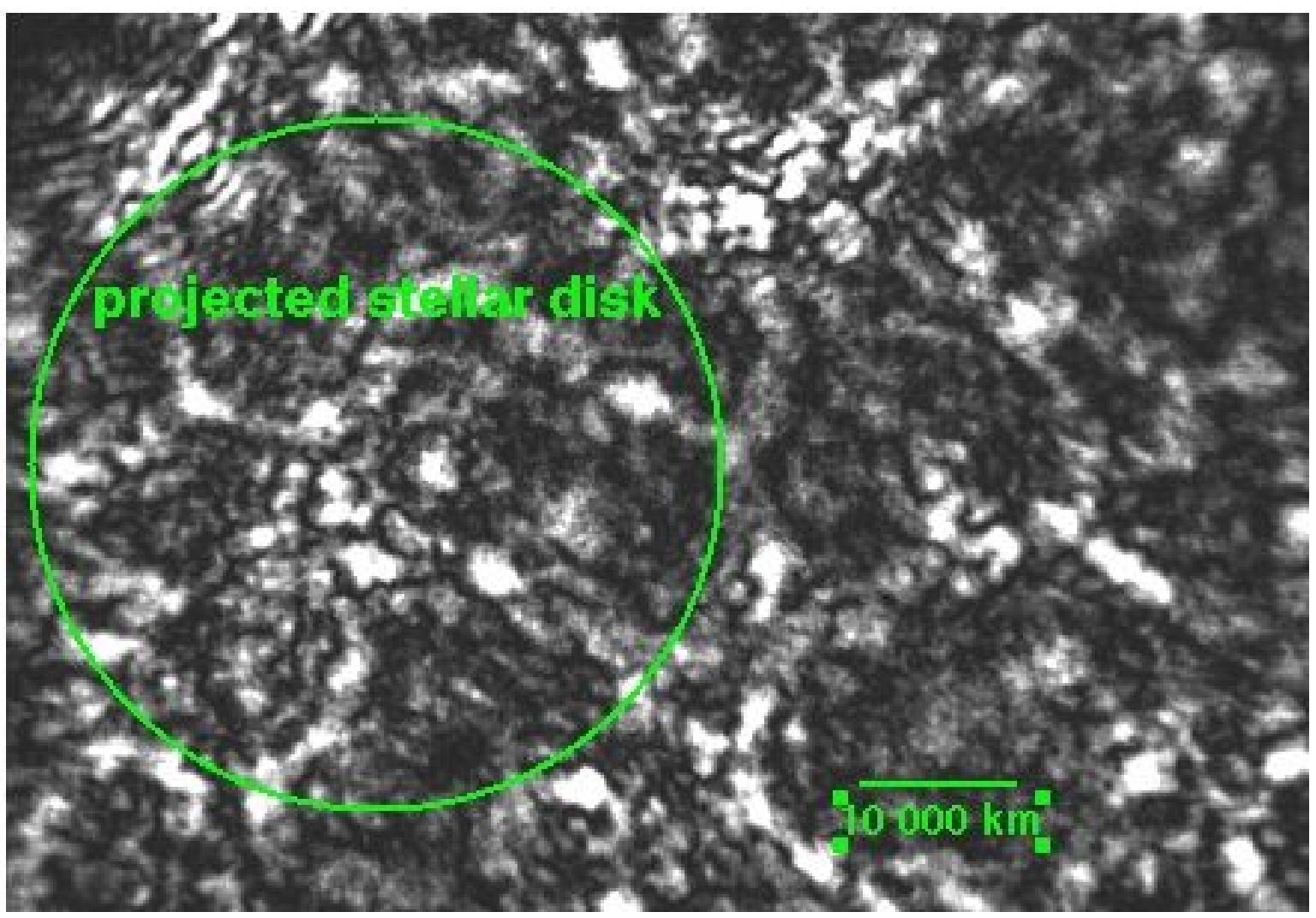}\\
\includegraphics[width=6cm]{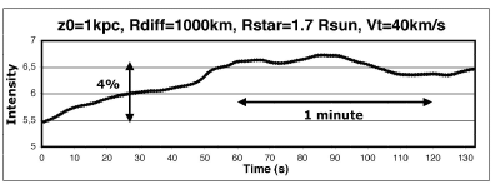}
}
\caption[]{\it
Phase delay $\Phi(x_1,y_1)$ after crossing a
turbulent cloud (left);
Illumination from a point source (up right);
Illumination from a A5V-LMC star versus time (down right).
\label{ecran}}
\end{figure}

\begin{figure}[h]
\centering
\parbox{8cm}{
\includegraphics[width=6.5cm]{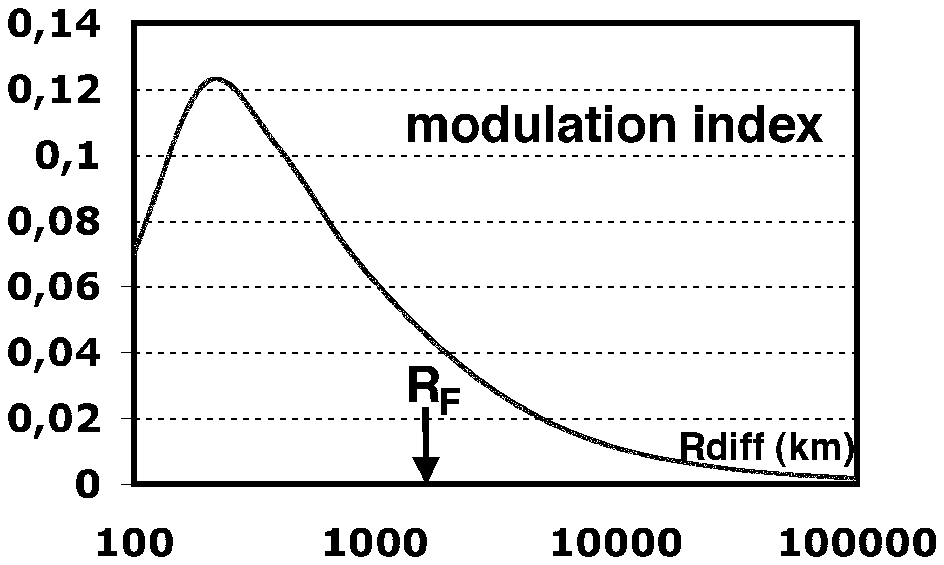}
}
\parbox{4cm}{\caption[] 
{\it
Modulation index of the intensity from a A5V LMC star,
as seen through a Galactic molecular cloud located at 1kpc
as a function of $R_{diff}$.}
\label{indice}}
\end{figure}

We have simulated the phase delay function
$\Phi(x_1,y_1)=2\pi\delta(x_1,y_1)/\lambda$ 
of a fractal cloud, described by the Kolmogorov turbulence law
(Fig. \ref{ecran} left). We calculated the diffraction image
of a {\it point-like} source on Earth (Fig. \ref{ecran} up-right)
from the Fast Fourier Transform of function
$exp\left[i \Phi(x_1,y_1)+i\frac{x_1^2+y_1^2}{2 R_F^2} \right].$
The illumination on Earth due to an {\it extended} incoherent source
is then obtained by
integrating the {\it intensities} due to each elementary source.
This is equivalent to integrate the intensity image of a point
source within the projected source stellar-disk of radius 
$r_s\times z_0/z_1$. The expected light-curve from a
A5V LMC star, as seen through a
cloud located at 1kpc, with $R_{diff}=1000 km$
and transverse speed $V_T=40 km/s$, is given in Fig. \ref{ecran} down-right.
Using this simulation, we have been able to estimate the modulation
index as a function of the crucial parameter $R_{diff}$ (Fig. \ref{indice}).

\section{Toward an experimental setup}
Table \ref{configurations} shows that the search for
diffractive scintillation
induced by a Galactic molecular cloud
needs the capability to sample
every $\sim 10\,\mathrm{s}$ (or faster) the luminosity of stars
with $M_V>20$ (A5V star in LMC or B0V in M31 --or smaller--),
with a precision of $\sim 1\%$.
This performance can be achieved using a telescope diameter
larger than two meters, with a
high quantum efficiency detector allowing a negligible dead-time
between exposures (like frame-transfer CCDs).
Multi-wavelength detection capability is highly desirable to
exploit the dependence of the diffractive scintillation pattern with
the wavelength.

The $1\%$ surface filling factor predicted if the galactic dark matter
is completely made of
gaseous structures 
is the maximum optical depth $\tau_{scint}^{max}$ for {\it all} the
possible 
scintillation regimes.
Let $\alpha$ be the fraction of halo made of such gaseous objects;
under the pessimistic hypothesis that strong diffractive regime occurs
only when a gaseous structure enters or leaves the line of sight,
the duration for this regime is
$\sim 5$ minutes (time to cross a few fringes)
over a total crossing time of $\sim 400$ days.
Then the diffractive regime optical depth should be at least
$\tau_{scint}^{min}\sim 10^{-7}\alpha$
and the average exposure
needed to observe one event of $\sim 5$ minute duration
is $10^{6}/\alpha\,\mathrm{star\times hr}$ \footnote{
Turbulence or any process creating filaments,
cells, bubbles or fluffy structures should decrease this estimate.}.
At isophot $M_V=23 mag/arcsec^2$
of LMC, SMC or M31, about $10^5$ stars per square degree with
$20<M_V<21$ --i.e. small enough-- can be monitored (\cite{Elson}; \cite{Hardy}).
It follows that a wide field detector is necessary
to monitor enough small stars.
\section{Foreground effects, background to the signal}
{\bf Atmospheric effects:}
Surprisingly, atmospheric intensity scintillation is negligible
through a large telescope ($m_{scint}\ll 1\%$ for
a $>1\,$m diameter telescope (\cite{dravins})).
Any other long time scale atmospheric effect such as
absorption variations at the sub-minute scale
(due to cirruses for example) should be easy to recognize as long
as nearby stars are monitored together.\\
{\bf The solar neighbourhood:}
Overdensities at $10\,\mathrm{pc}$ could produce a signal very similar
to the one expected from the Galactic clouds.
In such a case, even big stars would undergo
a strong diffractive scintillation,
contrary to the expectation from more distant screens;
simultaneous monitoring of various types of stars 
should allow one to discriminate effects due to nearby
gas or to remote gaseous structures.\\
{\bf Sources of background?}
Asterosismology, granularity of the
stellar surface, spots or eruptions
produce variations of different amplitudes and time scales.
A rare type of recurrent variable stars exhibit emission
variations at the minute time scale (\cite{sterken}),
but they are easy to identify from spectrum.
\section{Preliminary studies with the NTT}
Two nights of observations with the NTT were attributed
for testing the concept of scintillation in june 2004. We
monitored stars located behind or on the edge of Bok globules,
searching for scintillation signal due to the gas.
We got 5400 exposures of $T_{exp}=7s$ taken with the infra-red
SOFI detector, the only one operating with a short
readout time (5.5s).
The K-band is not
optimal for our purpose, but allows to detect stars through
the dusty Bok clouds. Unfortunately, only
948 measurements had a seeing smaller than 1.5 arcsec towards B68
globule. We produced the light curves of 2873 stars using the EROS
software. After selection and identification of known
artifacts (hot pixels, dead zones, bright egrets...), we found
only four stars with significant variability, all of four
interestingly located near the Bok globule limit.
As these limits are natural places for discontinuities,
there is a strong interest to clarify and confirm the status
of these stars with better quality data.
Nevertheless, we can already conclude
that the signal we are searching for should not be overwhelmed by
background.
\section{Conclusions and perspectives}
The opportunity to search for scintillation results from
the subtle coincidence between the arm-lever of
interference patterns due to hypothetic diffusive
objects in the Milky-Way and the size of the extra-galactic stars.
The hardware and software techniques required for such searches
are available just now.
If a signal is found, one will have
to consider an ambitious project involving synchronized telescopes,
a few thousand kilometers apart.
Such a project would allow to temporally
and spatially sample an interference pattern, unambiguously providing
the diffusion length scale $R_{diff}$, the speed and
the dynamics of the scattering medium.


\begin{thebibliography}{}
\bibitem[Afonso {\it et al.} 2003]{SMC5ans}
Afonso, C., {\it et al.} (EROS coll.), A\&A 400, 951 (2003)
\bibitem[Alcock {\it et al.} 2000]{MACHO}
Alcock, C., {\it et al.} (MACHO coll.), ApJ 542, 281 (2000)
\bibitem[De Paolis {\it et al.} 1995]{Jetzer1}
De Paolis, F. {\it et al.}, PRL 74, 14 (1995)
\bibitem[Dravins {\it et al.} 1997-98]{dravins}
Dravins, D. {\it et al.} {\em Pub. of the Ast. Soc. of the Pacific}
{\bf 109} (I, II) (1997), {\bf 110} (III) (1998)
\bibitem[Elson {\it et al.} 1997]{Elson}
Elson, R.A.W., Gilmore, G.F., \& Santiago, B.X. 1997,
{\tt astro-ph/9705149}
\bibitem[Hardy 1978]{Hardy}
Hardy, 1978, Pub. of the Astron. Soc. of the Pacific 90, 132
\bibitem[2000]{notenoughmachos}
Lasserre, T., {\it et al.} (EROS coll.), A\&A L39, 355 (2000)
\bibitem[Lyne \& Smith 1990]{lyne}
Lyne, A.G. \& F. Graham-Smith, {\it Pulsar Astronomy},
Cambridge University Press (1998)
\bibitem[Moniez 2003]{Moniez}
Moniez, M., A\&A 412, 105 (2003)
\bibitem[Pfenniger \& Combes 1994]{fractal}
Pfenniger, D. \& Combes, F., A\&A 285, 94 (1994)
\bibitem[Sterken \& Jaschek 1996]{sterken}
Sterken, C. \& Jaschek, C. {\it light Curves of
Variable Stars, a pictorial Atlas}, Cambridge University Press (1996)
\end{thebibliography}
\end{document}